\documentclass[pre,preprint,numerical]{revtex4-1}
\usepackage{graphicx}
\usepackage{amsmath,amssymb}
\usepackage{cprotect}
\usepackage{color}

\newcommand{\be}{\begin{equation}}              
\newcommand{\ee}[1]{\label{#1} \end{equation}}  

\begin{document}
\title{Waves in Strongly Nonlinear  Gardner-like Equations on a Lattice}
\author{Philip Rosenau} 
\affiliation{School of Mathematics,Tel-Aviv University,Tel-Aviv 69978, Israel }
\author{Arkady Pikovsky}
\affiliation{Institute of Physics and Astronomy, University of Potsdam,
Karl-Liebknecht-Str. 24/25, 14476 Potsdam-Golm, Germany}
\affiliation{National Research University,Higher School of Economics, \\
             25/12 Bolshaya Pecherskaya Ulitsa, 603155 Nizhny Novgorod, Russia}

\date{\today}

\begin{abstract}
We introduce and study a family of lattice equations which  may be viewed either as a strongly nonlinear discrete extension of the Gardner equation, or a non-convex variant of the Lotka-Volterra chain. Their deceptively simple form supports a very rich family of complex solitary patterns. Some of these patterns are also found in the quasi-continuum rendition, but the more intriguing ones, like interlaced pairs of solitary waves, or waves which  may reverse  their direction either spontaneously or due a collision, are an intrinsic feature of the discrete realm. 
\end{abstract}

\maketitle

\begin{quotation}
\end{quotation}

\section{Introduction}
With a progressing realization of their importance, strongly nonlinear lattices started to attract in the recent decades  an ever growing attention~\cite{Flach-95,Dey_etal-01,%
Rosenau-Pikovsky-05,Rosenau-Schochet-05,Rosenau-Schochet-05a,%
Pikovsky-Rosenau-06,HN-07,Ahnert-Pikovsky-09,%
Rosenau-Pikovsky-14,Rosenau-Zilburg-15,Zilburg-Rosenau-16,James-18}. This may be seen as a natural scientific evolution following in the foot steps of what may arguably be considered as a dawn of a non-linear science: the Fermi-Pasta-Ulam problem which, as envisioned by J. von Neuman~\cite{Lax}, used for the first time the newly born computer to carry a Gedankenexperiment, to test the long standing hypothesis of  energy equipartition of modes in solids due to  non-linearity. Though, given the primordial state of affairs, the studied chain, as solids model, was relatively very simple and short, their results  after few years in hibernation have led to the birth of the soliton and the chaos theories and, as computational facilities have advanced and spread around the globe, to ever more evolved theories and applications. Chains of coupled elements were introduced and used to simulate more realistic molecular interactions, micro-mechanical arrays, electrical transmission lines, optical lattices or photonic crystals, to name a few. These are well known facts which need no further elaboration, a succinct overview is presented in Ref.~\cite{Campbell-Flach_Kivshar-04}, and references therein.

A common feature of the aforementioned systems is the presence of a weakly nonlinear regime wherein the dynamics is determined by both linear and nonlinear parts. For such states
one may employ simple nonlinear expressions, though no one 
thought in earnest that  the quadratic force assumed by FPU could, apart of very small excitations, reproduce realistic molecular interactions. 
In fact, the numerical experiments carried a few years later have very clearly revealed that as the employed nonlinearity gained in strength, the idyllic recovery of the initial excitation by the FPU had to be replaced by a far more complex structures. If 
further arguments were ever needed, those results have clearly pointed to the 
necessity to probe deeper into the nonlinear regime. In 
the spring-mass 
context this amounts to probing the truly anharmonic domain which in its
ultimate rendition results in a system without linear waves (the so-called sonic vacuum), with all perturbations of the ground state 
being nonlinear  as well. Differently stated, a true  understanding  of nonlinearity's impact necessitates  to engage the nonlinear regime in its full glory, rather then as an  extension of its weakly nonlinear limit. In an analogy, to the extent that the infra-red and ultra-violet regime  represent opposite ends of the visible spectrum, similarly  the linear/weakly-linear  and the genuinely nonlinear regimes may be looked upon  as conceptual opposites, each of which should be  addressed on its own terms. As a typical signature of a genuinely nonlinear system one may consider the absence of a  non-scalable intrinsically small parameter to serve as a reference for a weakly nonlinear dynamics.

 Let us recall that aside of the conceptual importance of 
 the genuinely nonlinear limit, many systems by their very nature 
 are ab initio essentially nonlinear. The Hertz-like interaction 
 of elastic beads is perhaps the simplest mechanical realization 
 of such  system~\cite{Flach-95,Dey_etal-01,Flach-Gorbach-08,Sen-Hong-Bang-Avalos-Doney-08}. And whereas in mechanics one typically
deals with lattices described by coupled second-order differential equations, strongly nonlinear systems
have been also explored within the framework  of amplitude equations given via discrete Schr\"odinger-type    equations which are first-order in time ~\cite{Rosenau-Schochet-05,Rosenau-Pikovsky-14,James-18}.

In yet another set up, closer  to our present studies, a genuinely nonlinear conservative system emerges in a chain of coupled dissipative
oscillators with a limit cycle. In a strongly dissipative regime, where
the amplitudes of oscillators are truly stable, one may neglect
their variations and derive a closed conservative system for their phases only~\cite{Rosenau-Pikovsky-05,Pikovsky-Rosenau-06}. 
Such systems fall under what could be referred to as a Kuramoto type set-ups and  in a clear distinction from the Newton-like  chains, are systems of \textit{first order in time}. In a similar vein the Lotka-Volterra chain~\cite{Zilburg-Rosenau-16}  which originates in prey-predator system is  
another first order in time strongly nonlinear lattice. 

The problems to be addressed here are a direct off spring of the Kuramoto family
of strongly nonlinear first-order in time lattices~\cite{Rosenau-Pikovsky-05,Pikovsky-Rosenau-06,Rosenau-Pikovsky-20}, but instead of considering lattices of phase variables, with $2\pi$-periodic nonlinearities, we focus on lattices with polynomial nonlinearities.
For the present purpose  we shall restrict our attention to nonlinearities which are 
 sum of two monomials of opposite signs. In this form they are a discrete analog/extension
of the Gardner equation, which is a mixture of the K-dV and the mK-dV equations ~\cite{Slunyaev-01,GPPS-02,Kamchatnov-etal-12}. The lattice we address may be viewed as a simplified version of a generic phase lattice, allowing us to cope with some of the difficulties we have encountered there, and sure enough similarly to their original antecedents did not fail to produce new and surprising features which we believe are of independent interest.

Paper's plan is as follows. In Section~\ref{sec:bm} we introduce the basic model. In Section~\ref{sec:qc}
we present an analytical treatment of the traveling waves akin to the quasi-continuous, QC, rendition. In Sections~\ref{sec:btw} and \ref{sec:tptw}
we return to the full lattice set-up and explore  numerically simple and interlaced travelling waves.  
Simulations of the traveling solitary waves in a lattice are presented in Section~\ref{sec:dns} and reveal a number of fascinating phenomena not seen in the QC. We conclude with a
discussion in Section~\ref{sec:con}.

\section{The Basic Model}
\label{sec:bm}
Consider a genuinely nonlinear  conservative lattice with nearest-neighbor interactions:
\begin{equation}
\frac{du_k}{dt} = F(u_{k+1})-F(u_{k-1}), ~~k=\ldots,-1,0,1,2,\ldots \;
\label{eq:gl}
\end{equation}
where $F(u)$ is a smooth nonlinear function, to be specified in Eqs.~(\ref{eq:Fl}-\ref{eq:gl35})
below. To put the addressed problems in a perspective we note that our studies were motivated
by chains of conservatively coupled self-sustained (autonomous) oscillators wherein the state variable $u$ is the phase difference between neighboring oscillators reducing in the simplest 
case~\cite{Rosenau-Pikovsky-05,Pikovsky-Rosenau-06,Ahnert-Pikovsky-08}
 to $F(u)=\cos(u)$. With that particular choice the
lattice  \eqref{eq:gl} becomes genuinely nonlinear which is to say that there are linear waves with the resulting 
 solitary waves, the compactons and the kovatons, being almost compact (they decay at a doubly exponential rate) and essentially nonlinear entities. Notably, the model
\eqref{eq:gl} has been recently successfully used to describe  complex states
in a network of nano-electro-mechanical oscillators~\cite{Matheny_etal-19}. 

 In a previous work~\cite{Rosenau-Pikovsky-20} using a more general setting for 
phase waves in a chain of autonomous oscillators, it was assumed that 
$F(u)=\sin(\alpha - u)$. Though numerical simulations posed no particular 
difficulty, their direct 
analysis turned to be challenging forcing us   
to turn to their quasi-continuous rendition
and the resulting partial differential equation. Yet even there  we had to further restrict our self to a small amplitude regime which turned to be given by the Gardner equation which surprisingly enough provided a remarkably good qualitative description of the $\alpha\leq \pi/2$ domain, though elsewhere it was helpless. Since the periodic nature of $F(u)$ seemed to be the source of the encountered difficulties, we have adopted a simpler set-up of a 
polynomial $F(u)$, which replicate two of the vital singular transitions in the periodic case.  

 In passing we also note that in a lattice of a finite length $N$ and open boundaries,  system \eqref{eq:gl} is Hamiltonian  (the proof, due to H. Dullin,  follows Ref.~\cite{Pikovsky-Rosenau-06}). Since  for odd $N$ system \eqref{eq:gl} has an additional integral $K=\sum_{i=1}^{(N+1)/2} u_{2i-1}$, defining $Q(u)=\int F(u') du'$, the Hamiltonian of the lattice reads
\begin{equation}
    H(q_i,p_i)=\begin{cases} Q(q_1)+\sum_{i=1}^{m} Q(p_i)+\sum_{i=1}^{m-1} Q(q_{i+1}-q_i)+Q(K-s_m)& N=2m+1\;,\\
    Q(q_1)+\sum_{i=1}^{m} Q(p_i)+\sum_{i=1}^{m-1} Q(q_{i+1}-q_i)& N=2m\;.
    \end{cases}
\label{eq:ham1}    
\end{equation}
The used canonical variables are related to $u_i$ via $p_i=u_{2i}$, $q_i=\sum_{j=1}^i u_{2j-1}$. In the original variables the conserved energy can be thus expressed as $E=\sum_i Q(u_i)$.

Returning to our specific problem, we assume
a strongly nonlinear Gardner-type lattice as
\begin{equation}
 F(u) = mu^{n} -nu^{m} , ~~~1<m<n,
\label{eq:Fl}
\end{equation}
with integer $m,n$
(the values of the coefficients which are arbitrary and were chosen to simplify the analysis).
 Two typical cases will be analyzed:
\begin{equation}
\text{G23:}\quad \frac{du_k}{dt}=2u_{k+1}^3-3 u_{k+1}^2-2u_{k-1}^3+3 u_{k-1}^2,\qquad F(u)=2 u^3-3 u^2\;,
\label{eq:gl23}
\end{equation}
and
\begin{equation}
\text{G35:}\quad\frac{du_k}{dt}=3u_{k+1}^5-5 u_{k+1}^3-3u_{k-1}^5+5 u_{k-1}^3,\qquad F(u)=3 u^5-5 u^3\;.
\label{eq:gl35}
\end{equation} 
The G23 model is a direct replica of the original Gardner equation. The G35 model has different 
symmetries, with its resulting properties being quite different from those of the G23.

At this point we pause to note that since 
$F(u)$ vanishes at $u^*=\left(\frac{n}{m}\right)^{1/(n-m)}$, any
sequence of zeros and $u^*$ forms a stationary solution
on the lattice. In particular, one can have stationary
`pulse' solutions $\ldots,0,u^*,u^*,\ldots,u^*,0,\ldots$ of an arbitrary length. While the construction of these solution is trivial, their stability properties are rather complex and depend on both the total length of the lattice and the boundary conditions at its ends. Since we focus on traveling entities we shall not pursue further these solutions.

In addition to the standard conservation features, the G23 model is invariant under
\begin{equation}
u_k\rightarrow 1-u_k,
\label{eq:Sgl23}
\end{equation}
whereas G35, and any Gmn if both $m$ and $n$ are odd, 
is invariant under 
\begin{equation}
u_k\rightarrow -u_k 
\label{eq:Sgl35}
\end{equation}
and under
\begin{equation}
u_k\rightarrow (-1)^{k}u_k ~~~\text{and}~~~ t\rightarrow -t,
\label{eq:ZSgl35}
\end{equation}
which generates from any given solution  an alternating sign solution, referred to as a staggering solution in Section~\ref{sec:sc} below. Note that whereas the first two invariant features extend to the quasi-continuum, the third (Eq.~\eqref{eq:ZSgl35}) is obviously an exclusive feature of the discrete realm.

We stress that the novel features of the presented problems are due to the  {\it non monotonic} nature of the assumed $F(u)$. A widely studied version 
with $F=\exp(u)$, also referred to as the ladder equation, is integrable (and equivalent
to the Toda lattice)~\cite{Manakov-75,KM-75}. More relevant to our 
problem is the discrete $K(\alpha,\alpha)$ equation with the monomial $F=\gamma u^{\alpha}$ (cf Ref. \cite{Rose-Zilburg}):
\begin{equation}
  \frac{du_k}{dt}=\gamma\big(u_{k+1}^\alpha-u_{k-1}^\alpha \big),  ~~~~\text{~where~~$\gamma$~is~a~const.} \label{eq:K(n,n)}
\end{equation}
which may be viewed as  a
limiting case of the generic lattice \eqref{eq:gl},\eqref{eq:Fl},
either for large $u$, where $\gamma=m$, $\alpha=n$, or for small $u$, where $\gamma=-n$, $\alpha=m$. 

 Similarly to our previous studies, Eqs. \eqref{eq:gl},\eqref{eq:Fl} have a deceptively simple appearance with their numerical integration being  straightforward, but the understanding of the underlying dynamics is a very different matter. For starters, our ability to analyze nonlinear discrete systems, or better yet, to unfold their coherent structures, is very limited, and far more 
 inferior to our ability to analyze systems represented by partial differential equations (PDEs). This naturally leads us to represent the discrete system via the Quasi-Continuum~\cite{Rosenau-03,Rose-Zilburg}, QC (Section~\ref{sec:qc}). Yet this approach has its difficulties, for whereas in weakly nonlinear systems the stationary solution provides a  reference level with respect to which one carries the asymptotic expansion, as aforementioned, the present problem lacks an essential small parameter (the distance between the discrete nodes may be scaled out). Therefore the adopted QC route is not asymptotic and its value can be {\it judged only by its utility}. Luckily, and one could say: very luckily, we find that as in a number of our previous works \cite{Rosenau-Schochet-05,Rosenau-Pikovsky-05,Rosenau-Schochet-05a, Pikovsky-Rosenau-06},  in certain regimes QC provides  an excellent approximation of its discrete antecedent, though elsewhere it turned to be of far more limited use.      

\section{Traveling Waves in a Quasi-Continuum}
\label{sec:qc}
\subsection{The Quasi-Continuum Rendition}


Replacing the finite differences of the 
 discrete lattice equations with a continuous spatial derivatives up to  a third order yields the QC rendition of the original problem (see Refs.~\cite{Rose-Zilburg} or \cite{Rosenau-03} for a detailed exposition of the desired QC approach and its limitations) of Eqs. \eqref{eq:gl},\eqref{eq:Fl}:
\begin{equation}
\frac{1}{2}\frac{\partial}{\partial 
 t}u = \frac{\partial}{\partial x}{\cal L}^{2}F(u) ~~~\text{where} ~~~{\cal L}^{2}=1 + \frac{1}{6}\frac{\partial^{2}}{\partial x^{2}} .
\label{eq:qcpde}
\end{equation}
Similarly to their discrete antecedents, we have the following conservation laws of Eq.~\eqref{eq:qcpde}:
\begin{equation}
I_{1}=\int{udx} ~~~\text{and} ~~~ I_{3}=\int{Q(u)dx}~~\text{where}~~Q(u)=\int^u {F(u')}du'\;, 
\label{eq:qccon}
\end{equation}
and an unusual Lagrangian structure
\begin{equation}
{\emph{L}}_{agrange} =\int\int\Big[\frac{1}{4}\psi_{x}\psi_{t}+ Q\Big({\cal L}\psi_{x}\Big)\Big]dxdt,
\label{eq:lag}
\end{equation}
where
$ u= {\cal L}\psi_{x}$,
which begets the conservation of the momentum $\int \psi_x^{2} dx$, and in the original variables
\begin{equation}
I_{2} =\int{ u{\cal L}^{-2}u}\;dx. 
 \label{eq:qccon3}
\end{equation}

\subsection{Solitary Waves in QC}
We now turn our attention to solitary waves $u(x,t)=u(s=x-\lambda t)$. Since $F(0)=0$, then upon one integration we have
\begin{equation}
{\frac{1}{2}\lambda}u + \left(1+\frac{1}{6}\frac{d^2}{ds^2}\right)F(u)=0,
 \label{eq:ode1}
\end{equation}
or
\begin{equation}
\frac{1}{2}{\lambda}u + F(u)+\frac{1}{6}\frac{d}{d s}F'(u)\frac{d u}{d s}=0.
 \label{eq:ode2}
\end{equation}
A crucial role is played by the zeros of $F'(u)$ wherein equation \eqref{eq:ode2} 
becomes singular. Also, since $F'(u)=mn(u^{n-1}-u^{m-1})$, it vanishes at both $u=0$ and $u=1$. 

  Multiplying \eqref{eq:ode2} with $F'(u) u_s$ and integrating once  yields the energy integral (since we are after solitary waves the integration constant was discarded) 
  \begin{equation}
F'^{2}(u)u_{s}^{2} + 6P(u) = 0  ~~~ \text{where}~~ P=  - \frac{\lambda}{nm}\Big(\frac{u^{m+1}}{1+m}-\frac{u^{n+1}}{1+n}\Big) + \Big(\frac{u^{m}}{m}-\frac{u^{n}}{n}\Big)^{2},
 \label{eq:Ene1}
 \end{equation}
and $P(u)$ is the potential. Cancelling $u^{2m-2}$ on both sides begets
 \begin{equation}
\frac{m^{2}n^{4}}{6}(1-u^{n-m})^{2}u_{s}^{2} - \Big(\frac{n\lambda}{m(n+1)}\Big)\Big(\frac{n+1}{m+1}-{u^{n-m}}\Big)u^{3-m} + \Big(\frac{n}{m}- u^{n-m}\Big)^{2}u^{2}=0 \;.
\label{eq:Pmn}
\end{equation}
The singularity at $u=0$ which causes degeneracy of the highest order operator and a local loss of uniqueness, allows us to construct
compactons - solitary solutions with a compact support. Indeed, since the uniqueness of
solutions of \eqref{eq:Ene1} is violated at $u=0$,
one may 'glue' there the nontrivial solution of \eqref{eq:Pmn} with the trivial $u=0$  solution. This  begets the compactons depicted in Figs.~\ref{fig:qc23},\ref{fig:qc35}.

Now, for $u=1$ to be an admissible solution we need  the 'total force'
in~\eqref{eq:ode2}
 $$f \doteq \frac{\lambda}{2}u +F$$
  to vanish there. This defines the critical velocity
\begin{equation}
\frac{\lambda_{1}}{2}=n-m\;.
 \label{eq:qclam1}
\end{equation}

Turning to Eq.~\eqref{eq:Pmn}, which is also singular at $u=1$, for $u=1$ to be admissible as a solution the potential has also to vanish {at this point}. This imposes an additional constraint  on the velocity
 \begin{equation}
\frac{\lambda_{2}}{2} = \frac{(n-m)(m+1)(n+1)}{2nm}\;.
 \label{eq:qclam2}
 \end{equation}
 Consistency demands that $\lambda_{1}=\lambda_{2}$, which constrains the admissible powers in $F(u)$:
  \begin{equation}
  n=\frac{m+1}{m-1}\;.
  \label{eq:n-m}
  \end{equation}
 Thus, since $1<m<n$,  $(m,n)=(2,3)$ emerge as {the  only pair of integers} for which the constraint  \eqref{eq:n-m} holds and thus supports formation of solutions incorporating singularities at both $u=0$ and $u=1$
 (the so-called kovatons~\cite{Rosenau-Pikovsky-05,Pikovsky-Rosenau-06}).

In fact, since $P' =F' f$ and $P'' = F''f+ F'f'$, then at $u=1$ where $F'=0$ and $P' =0$, for $P''$ to vanish as well, we need that both $f$ and $P$ vanish there with the same velocity. When these conditions are satisfied, $u_{s}$ vanishes, and both compact kink/anti-kink and $u=1$ become admissible solutions, which underlines the formation of kovatons (their explicit construction will be provided shortly). Formally,
$$\left. -u_{s}^{2}\right|_{u=1} = \left. \frac{P(u)}{F'^{2}(u)}\right|_{u=1} =\left. \frac{P'(u)=F'f}{2F'F''}\right|_{u=1}= \left. \frac{f}{2F''}\right|_{u=1}.$$
Since $F'' =-nm(n-m)\neq 0$ at $u=1$, therefore if both $f$ and $P$ vanish at the same velocity, $u_{s}$ will vanish as well.

On the other hand, if the potential and the force vanish at $u=1$ at  different speeds, then $u_{s}\neq 0$,  $u=1$ is not a solution and kovatons cannot form.

The existence of kovatons may  be embedded into a bit more general framework via the invariance of the equations of motion, whether discrete or QC, under
 \begin{equation}
u\rightarrow 1-u ~~\text{and}~~F(u)\rightarrow -F(1-u)+const., 
 \label{eq:Inv}
 \end{equation}
or simply into the condition $F'(u)=F'(1-u)$. If $F$ 
is a  polynomial of degree $n$, one may deduce the constraints for this condition to hold. For $n=3$ we obtain the G23 model. Among quintic choices the used $F(u)$ in the G35 model provides a counterexample whereas $F'=u^{2}(1-u)^{2}$  supports kovatons. A more general class admitting kovatons is afforded by $F'= g(u)g(1-u)$, $g(u=0)=0$, where 
$g$ may be any smooth function, or by $F=\sin^{2n}(\pi u)$, etc. 


\subsection{The G23 case} 
\label{sec:qcg23}

\begin{figure}[!htb]
\includegraphics[width=\textwidth]{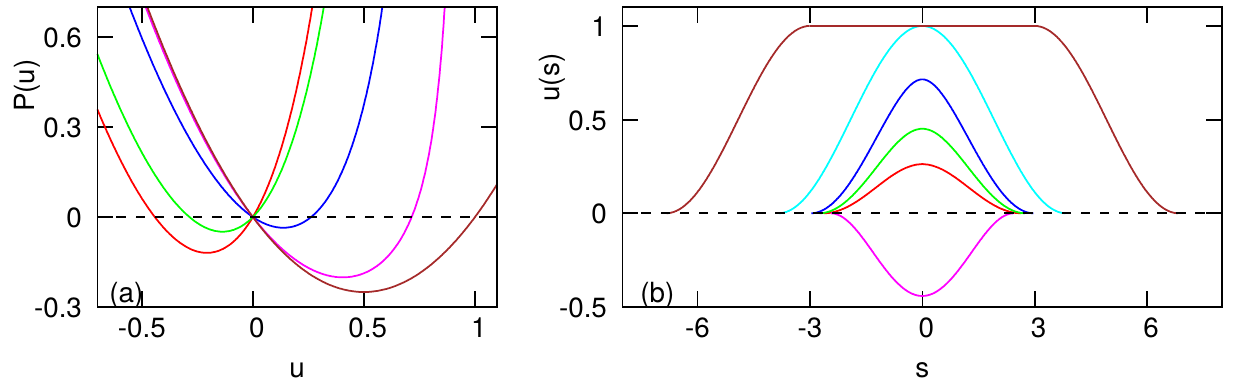}
\cprotect\caption{Panel (a): The effective potential $P_{23}(u)/(1-u)^2$ for six values of $\lambda's$ (from left to right: $\lambda=-2.5,-1.5,1, 1.9,2$) Note that only in the last case the effective potential remains bounded. 
Panel (b): 
compacton profiles for $\lambda=-2.5, 1,1.5,1.9$ (from bottom to top); the basic kovaton \eqref{eq:kov1},
and kovaton \eqref{eq:kov2} with $s_0=3$.
}
\label{fig:qc23}
\end{figure}

Let $m=2$ and $n=3$. After  cancelling of the common $u^{2}$ factor, Eq. \eqref{eq:Pmn}  reads
 \begin{equation}
\frac{3}{2}(1-u)^{2}u_{s}^{2} + P_{23}(u) = 0  ~~~ \text{where}~~ P_{23}(u)=  -\frac{\lambda}{2} u + \frac{3}{8}(6+\lambda)u^{2}-3u^{3}+u^{4}.
 \label{eq:P23}
 \end{equation}
 For $\lambda<2$, Eq.~\eqref{eq:P23}  enables to calculate numerically the shape of the solitary traveling wave (compacton) and, due to the singularity at $u=0$, match  
  the periodic solution with the trivial
 state. Several such solutions are displayed in Fig.~\ref{fig:qc23}.
 
 For large and small amplitudes we may derive explicit expressions for the solitary waves.
 For small amplitudes, neglecting the cubic part in the equation of motion, the resulting compacton takes a simple form 
  \begin{equation}
  u=\frac{2\lambda}{9}\cos^{2}\left(\frac{\sqrt{6}s}{4}\right)H\Big({2\pi}-{\sqrt 6}|s|\Big);,
   \label{eq:com1}
   \end{equation}
   {where $H(\cdot)$ is the Heaviside function.}
The invariance under $u \to 1-u$, begets also  compact drops hanging from the $u=1$ ``ceiling''. Assuming $0\leq v=1-u$ to be small, we have
  \begin{equation}
   u=1-\frac{2\lambda}{9}\cos^{2}\left(\frac{\sqrt{6}s}{4}\right)H\Big({2\pi}-{\sqrt 6}|s|\Big).
    \label{eq:g23com2}
 \end{equation}
 
In the limiting $\lambda=2$ case  $P_{23}(1)=P'_{23}(1)=P''_{23}(1)=0$,  the problem simplifies
\begin{equation}
(1-u)^{2}\Big[3u_{s}^{2} - 2u(1-u)\Big] = 0~~~\text{where} ~~ s=x-2t.
 \label{eq:kov}
\end{equation}
Clearly,  the singularity at $u=1$ is now accessible and we obtain a kink and/or anti-kink of {\it a finite span}. Tied together back-to-back they form the basic
kovaton (see Fig.~\ref{fig:qc23})
\begin{equation}
u=\cos^{2}\left(\frac{s}{\sqrt{6}}\right)H\Big(\sqrt{\frac{3}{2}}\pi-|s|\Big).
\label{eq:kov1}
 \end{equation}
Better yet, since $u=1$ is now solution as well, we may form a combined three-some entity, a flat hat
kovaton (see the right panel of Fig.~\ref{fig:qc23}):
\begin{equation}
u_{kov}(s) = \left\{\begin{array}{cc}
A & \mbox{~~~for $-\sqrt{\frac{3}{2}}\pi\leq s+s_{0}\leq 0$} \\
1 & \mbox{for $|s|\leq s_{0}$} \\
A & \mbox{~~~for $0\leq s -s_{0}\leq \sqrt{\frac{3}{2}}\pi$}
\end {array}
\right.
\label{eq:kov2}
\end{equation}
 where $A=\cos^{2}\Big(\frac{s_{0}-|s|}{\sqrt{6}}\Big)H\Big(\sqrt{\frac {3}{2}}\pi+s_{0}-|s|\Big)$ and $0<s_{0}$ is  an arbitrarily chosen constant,  determining top's width. Again, due to problem's  invariance under $u\rightarrow 1-u$,
  we also have an anti-kovaton
\begin{equation}
 u_{anti-kov}=1-u_{kov}.
  \label{eq:antikov}
 \end{equation}
 
  Comparing basic kovaton's \eqref{eq:kov1}  width with its small amplitude sibling in \eqref{eq:com1}, we notice that the kovaton is  $\sim 50$ percent wider. Compacton's widening with the amplitude indicates that the dispersion-convection balance tilts with amplitude toward the dispersion.
  
  For a later use we record the regime of large amplitude waves, wherein $F$ has already changed its sign and its quadratic part may be ignored, leaving us with only the cubic piece. In the resulting problem 
  \[
  \frac{du_k}{dt}=2(u_{k+1}^3-u_{k-1}^3)\;,
  \]
the solitary waves propagate to the left ($\lambda<0$) and their shape is easily derived via the corresponding QC equation
\begin{equation}
u(s)=\pm\left(\frac{3|\lambda|}{8}\right)^{1/2}
\cos\big(\sqrt{\frac{2}{3}} s_{+}\big)H\big(\pi-\sqrt{\frac{8}{3}} |s_{+}|\big) ~~~\text{where}~~s_{+}=x+|\lambda|t\;.
\label{eq:p3c}    
\end{equation}
The $\pm$ sign expresses the reduced equation's invariance under $u_{k}\rightarrow - u_{k}$.

\subsection{The G35 case}
\label{sec:qc35}

\begin{figure}[!htb]
\includegraphics[width=\textwidth]{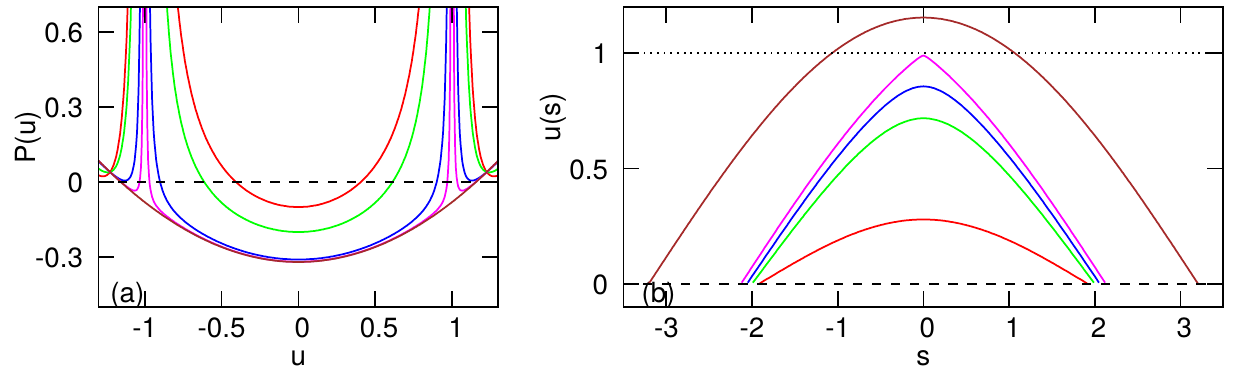}
\cprotect\caption{Panel (a): The effective potential $P_{35}(u)/(1-u^2)^2$ for different values of $\lambda$ (from top to bottom: $\lambda=1,2,3,1,3.199,3.2$). Note that the effective potential remains finite only at the limiting velocity $\lambda=32/5$,  thus enabling the corresponding solution's trajectory to cross it. Panel (b):compacton profiles for $\lambda=0.5, 2.5,3,3.199, 3.2$ (from bottom to top). Note the support of  the limiting  solution which is much wider than its velocity-wise close neighbors. 
}
\label{fig:qc35}
\end{figure}

We now consider the $m=3$ and $n=5$ case.
 Following the reduction by a common $u^{4}$ factor in \eqref{eq:Pmn}, we have 
\begin{equation}
(1-u^{2})^{2}u_{s}^{2} + P_{35}(u) = 0  ~~~ \text{where}~~ P_{35}(u)=  -\frac{\lambda }{10}(1- \frac{2}{3}u^{2})+ \frac{2u^{2}}{3}(1-\frac{3}{5}u^{2})^{2}\;.
\label{eq:p35}
 \end{equation}
 Its compact solutions are displayed in Fig.~\ref{fig:qc35}.
As in the previous case the singularity at $u=0$, and the associated local loss of uniqueness, enable to glue the periodic solution with the trivial state to form a compacton which though continuous has its first derivative jump at $u=0$.

As in the other case, for small amplitudes one keeps only the lower order part in $F(u)$ with the  explicit form of compactons, up to a normalization, given via \eqref{eq:p3c}.
 In the opposite case of solitary waves with large amplitudes, keeping only the quintic part we have 
\begin{equation}
u(s)=\pm\left(\frac{5|\lambda|}{18}\right)^{1/4}
\cos^{1/2}\Big(\frac{2\sqrt{6}}{5} s_{+}\Big) H(2\pi\sqrt{6}-10|s_{+}|)~~~\text{where}~~ s_{+}= x+|\lambda|t\;.
\label{eq:p5c}    
\end{equation}
As in~\eqref{eq:p3c}, the resulting compactons propagate to left, $\lambda<0$.

We now address the affairs near the $u=1$ singularity .
From~\eqref{eq:p35} we have 
$P_{35}(u=1)=(16/5 -\lambda)/30$. At
the critical velocity $\lambda=16/5$, $P_{35}(1)=P'_{35}(1)=0$ 
(since $f\neq 0$, $P''_{35}(1)\neq 0$). At the limiting velocity we thus have
\begin{equation}
(1-u^{2})^{2}\Big[25u_{s}^{2} - 6(\frac{4}{3}-u^{2})\Big] = 0~~~\text{where} ~~s=x-\frac{16}{5}t\;.
\end{equation}
 Though in the present case {\it kovatons
 cannot form}, at the limiting velocity, and only at this velocity, the singularity at $u=1$ allows the regular solution trajectory to 'sneak through' and cross it (in  $u=1$ vicinity $u\sim 1 +as+bs^{2}+...$ where $a=\frac{\pm 1}{5\sqrt{3}}$ and $b$ is a constant) and  the limiting solution assumes  a simple form
\begin{equation}
u(x,0) = \frac{2}{\sqrt 3}\cos\Big(\frac{\sqrt{6}x}{5}\Big)H\Big(5\pi-2\sqrt{6}|x|\Big).
\label{eq:g35top}
\end{equation}
Notably, in addition to this exceptional solution, 
the
 singularity at $u=1$ admits also a  solution which is non-smooth at its top 
\begin{equation}
u(x,0) = \frac{2}{\sqrt 3}\cos\Big(\frac{\pi}{6}+\frac{\sqrt{6}|x|}{5}\Big)H\Big(5\pi-3\sqrt{6}|x|\Big),
\label{eq:peakon}
\end{equation}
 attained at $u=1$ and is thus a {\it compact peakon}: it has a finite support, is everywhere continuous, but its first derivative at the peak switches its sign $\pm 1/\sqrt 3$. From Fig.~\ref{fig:qc35} one also notes that whereas the exceptional solution's support  undergoes a sizeable jump with respect the the support of its speed-wise close neighbors, peakon's support  is a continuous extension of the support of its velocity-wise close neighbors and may thus be considered their natural extension.

\subsection{Relevance of a Quasi-Continuum based analysis}
Before leaving the QC realm we need to clarify the role of the QC in elucidating the discrete patterns, but first some basic facts. Whereas the singularity at $u=1$   bounds the domain accessible by the PDEs, as we shall shortly see,  the discrete antecedents have no such barrier and $u=1$ is merely a 'sign road' of things to change.  In fact, both the G23 and the G35  discrete problems have  large amplitude solutions which are far beyond the access of their respective QC PDEs renditions. Yet though those PDEs  failed to describe the dynamics beyond the $u=1$ barrier, the QC approach may be still  of use if applied  separately to the small and large 
amplitude domains, see Eqs.~\eqref{eq:K(n,n)}
and solutions~\eqref{eq:com1}, \eqref{eq:p3c},\eqref{eq:p5c}. There is no contradiction here because {\it the singularity is a barrier of the PDE's and not of the original problem}. We may bypass the 'mine field' at $u=1$ by  splitting the PDE representation into a 'sub-critical' domain,  valid up to the singularity, and a 'super-critical' QC description, applied  at large amplitudes,  Eq.~\eqref{eq:K(n,n)}, where both $F$ and  $F'$ have already changed their signs, with the corresponding  large amplitude solitary solutions  recorded in ~\eqref{eq:p3c} and ~\eqref{eq:p5c}, respectively.

And yet, though  the discrete problems appear formally to be oblivious of the singular barriers, nonetheless those barriers appear to be somehow implicitly imprinted in the system (similar phenomenon was also observed in \cite{Rosenau-Schochet-05a}). We shall find again and again that the more interesting  action in the discrete realm takes place in a close vicinity of those singular transitions unfolded by  PDEs which per se are no longer valid there! 

Existence of a thin layer where {\it the discrete effects are essential}, is analogous to the emergence of shock waves in an ideal gas in a boundary layer where, whenever Euler equations break down, one has to restore the viscosity, or better yet, to evoke the original gas-kinetic description. Yet away from the breakdown zone Euler equations work well.  Notably, in both gas dynamics and in our problem, the more complex and intriguing phenomena occur in the transition zone, marked by the ideal PDEs, yet described only via the original kinetic or the discrete set-up to be unfolded next.

\section{Basic Traveling Waves, TWs, on a Lattice.}
\label{sec:btw}
\subsection{Integral Formulation}
\label{sec:nie}
Returning to the original discrete problem we seek solitary traveling waves on the lattice~\eqref{eq:gl} of the standard form
\[
u_k(t)=U\left(t-{k}a\right)~~~\text{where}~~a=1/\lambda\;.
\]
Substitution in \eqref{eq:gl} yields a delayed-advanced equation
\[
\frac{dU}{dt}=F[U(t-a)]-F[U(t+a)]\;.
\]
Assuming that $U(\pm\infty)=0$, we integrate the last equation and set $t= as$ to simplify the resulting nonlinear integral equation
\begin{equation}
\lambda U(s)+\int_{-1}^1 F[U(s+s')]\; ds'=0\;
\label{eq:nint}
\end{equation}
which will be explored next.

\subsection{The Newton-Raphson Algorithm.}
Equation~\eqref{eq:nint} is in a form which enables to apply the Newton-Raphson algorithm using
the standard continuation approach~\cite{Doedel-07}. The free parameter $\lambda$,
was used to search for solution's branches.
 Assuming a  given 
 $(\lambda_0,U_0)$ solution, to find a new one we append
 Eq.~\eqref{eq:nint} with an auxiliary equation $$(\lambda-\lambda_0)^2 +(N[U]-N[U_0])^2=\Delta^2\;,~~~\text{where}~~ N[U]=\int U(s)ds$$
  is the norm of the 
solution, and $\Delta$ a (small) 
shift parameter.  A solution pair $(\lambda,U)$ of the extended system is then sought  via the Newton-Raphson algorithm.
In practice we have used $\Delta=10^{-2} - 10^{-3}$,
and the integral in \eqref{eq:nint} was evaluated using the Simpson formula, with a typical step of $~ \Delta s=0.02$. Solving the joint system for $U(s)$ and $\lambda$  we advance along the solutions branch.  Since the TW tails decay at a doubly-exponential rate~\cite{Dey_etal-01,Pikovsky-Rosenau-06,Ahnert-Pikovsky-09},   for all practical purpose their span may be considered finite. Therefore in choosing the integration domain   we take $|s|\leq L$; with $L$ assumed to be an integer and chosen such that $|U(L-1)|<10^{-16}$ (if this condition was violated the range was extended to $L\to L+1$).
Once solution was found (only symmetric solutions, $U(-s)=U(s)$, were explored), it was tested for stability  via a direct numerical
simulation of the original lattice equation on a small lattice with its size being twice the size
of the TW. If  after
$10^4$ rotations the solution remained unchanged, it was declared stable.

\subsection{Basic TW branches in the G23 model.}

\begin{figure}[!htb]
\includegraphics[width=\textwidth]{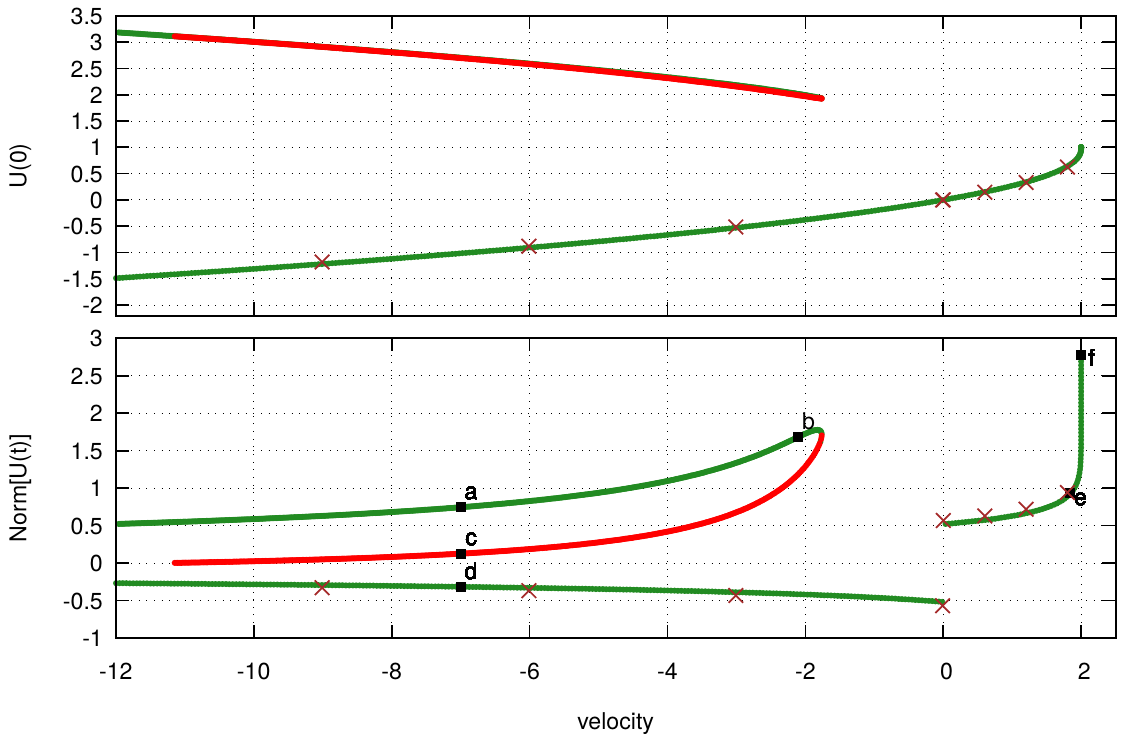}
\cprotect\caption{Basic traveling wave branches in the G23 model, Eq.~\eqref{eq:gl23}. Green: stable waves, red: unstable waves. Brown crosses: values predicted by the QC theory.
Note the green color which hides behind the red in the upper branch of the upper plate and the solutions $a$ and $c$ which have the same amplitude and velocity, but different norms (mass). }
\label{fig:bd_23}
\end{figure}

\begin{figure}[!htb]
\includegraphics[width=\textwidth]{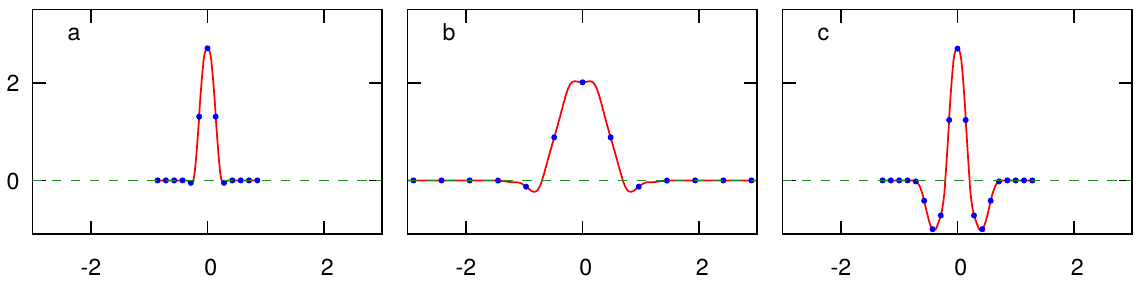}
\includegraphics[width=\textwidth]{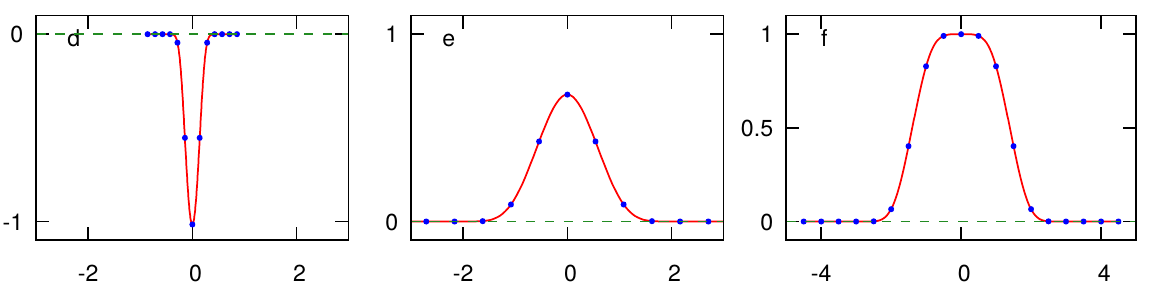}
\cprotect\caption{Basic traveling waves,TW, profiles $U(t)$ at the corresponding lettered points 
on the diagram in Fig.~\ref{fig:bd_23}. Red lines: profiles as function of time. Blue circles:
snapshots on the lattice.Note that TW velocity in cases (a), (c) and (d )is the same, though with case (c) being unstable, see Fig.(3), one may not be able to observe its propagation. }
\label{fig:bd_23f}
\end{figure}

In Fig.~\ref{fig:bd_23} we present the branches of the TWs found in the 
G23 lattice~\eqref{eq:gl23}. Their features may be summarized as follows:

\begin{itemize}
\item There is solutions branch  which is very well described by the QC, Sec.~\ref{sec:qcg23} (marked with
brown crosses in Fig.~\ref{fig:bd_23}). The displayed solutions are noted at lettered points d,e and f  in 
Fig.~\ref{fig:bd_23f}. {\it All waves along this branch are stable.}  For small amplitudes  Eq.~\eqref{eq:gl23} may be approximated via \eqref{eq:K(n,n)} with $\alpha=3$, 
Being   at this regime invariant under $u\to -u$, $t\to -t$,  implies that  small-amplitude compactons are nearly symmetric.

\item At large amplitudes wherein $F(u)$ has already changed its sign, the unfolded branch has {\it negative velocities}. The  ~\eqref{eq:p3c} may be viewed as its QC rendition. This branch relies on $F(u)$ being negative and cannot be extended  to small amplitudes/velocities. The  patterns corresponding to lettered points a,b and c are depicted in 
Fig.~\ref{fig:bd_23f}. Notably, only waves with a larger norm are stable (in the upper amplitude-velocity diagram the stable and the unstable branches nearly overlap, a much clearer distinction  between these branches  is provided by the lower norm-velocity 
panel). Note also that on the presented large amplitude branch, the quadratic part of $F$ though small is not completely negligible which affects the proximity between the analytical and the numerical results. It may also effect wave's stability, for unless the pulse is truly large, for a sizeable part of its profile $F(u)$ is positive with this part's tendency to propagate to the right.  $u$ has to be considerably over $F's$  transition value $3/2$ for the dynamics to enforce  a stable propagation to the left.  
\end{itemize}

\subsection{Basic TW branches in the G35 model.}

\begin{figure}[!htb]
\includegraphics[width=\textwidth]{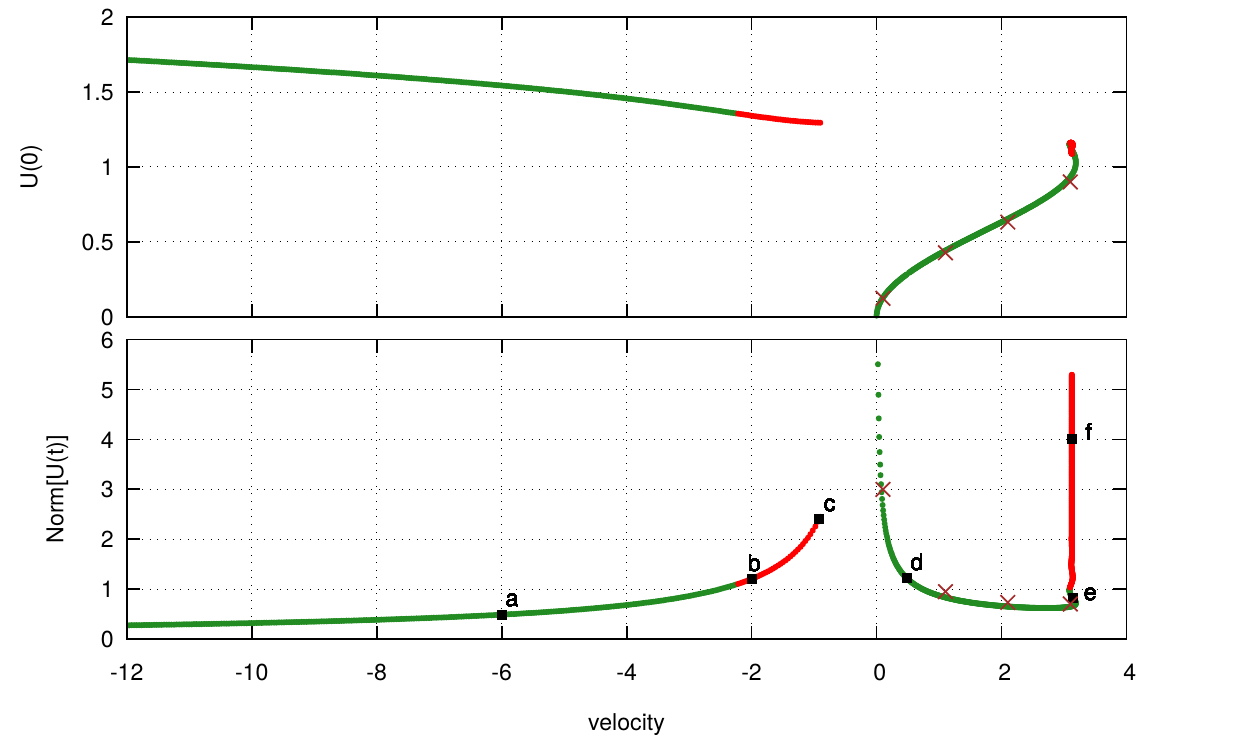}
\cprotect\caption{The basic TW branches in the G35 model, Eq.~\eqref{eq:gl35}.  Green: stable TWs, red: unstable TWs. Brown crosses show the corresponding values of TW in the QC rendition. The upper panel clearly shows an amplitude gap which may cause TW located close to the right edge of the left branch to 'hop' to the right branch, with a consequent switch of direction, whether due to inner instability or collision, if the resulting amplitude falls below the minimal admissible value of the left branch. Such  scenarios are displayed on Figs. \ref{fig:stpt35-1} and \ref{fig:coll35}. Fig. \ref{fig:coll35st} shows a reverse scenario: collision between waves on the right branch causes one wave to hop to the left branch and then hop back. And whereas depending on the direction of hoping wave's amplitude may increase or decrease, their mass hardly changes.}
\label{fig:bd_35}
\end{figure}

\begin{figure}[!htb]
\includegraphics[width=\textwidth]{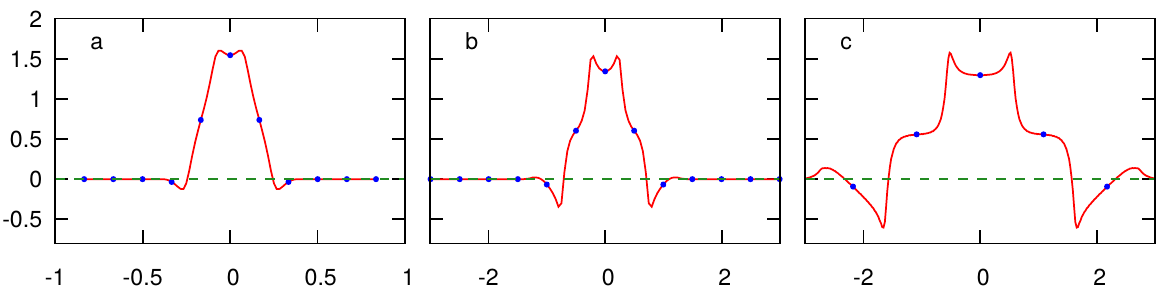}
\includegraphics[width=\textwidth]{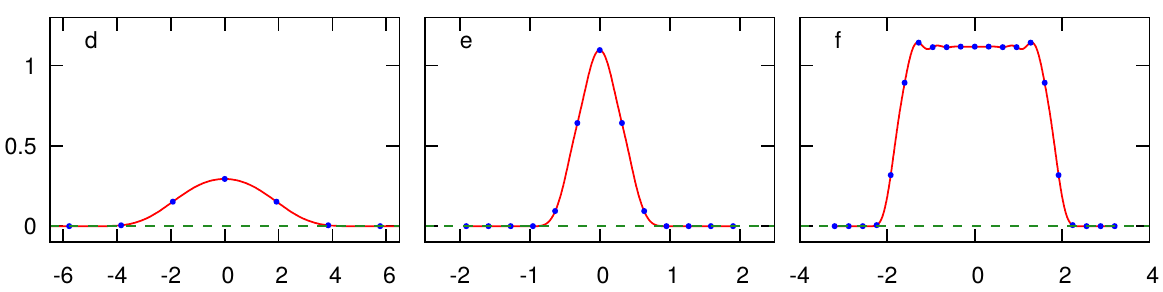}
\cprotect\caption{Display of the basic G35 TWs located at the corresponding lettered points in  Fig.~\ref{fig:bd_35}. Red lines: amplitude as a function of time. 
Blue circles:
amplitude snapshots on the lattice. Surprisingly enough the amplitude of the flat-like case (f) is not 
$\sqrt{5/3}$ where $F(u)$ vanishes, but much closer to $2/\sqrt{3}$, the amplitude of the only QC compacton solution, see Fig.~\ref{fig:qc35}, that sneaks through the $u=1$ barrier, though its velocity $~25/8$ is slightly under the maximal QC speed $16/5$ }
\label{fig:bd_35f}
\end{figure}

We now address the basic solitary TWs in the G35 model~\eqref{eq:gl35}.
Due to its $u\to -u$ symmetry, it suffices to display in Figs.~\ref{fig:bd_35},\ref{fig:bd_35f} only  
waves with positive amplitudes, and skip their symmetric counterparts with negative amplitudes. The solitary waves exhibit the following properties:
\begin{itemize}
    \item As in the G23 case, the G35 small amplitudes regime is very well described by the QC (Sec.~\ref{sec:qc35}). However, at larger amplitudes as one approaches the transition zone the features of the found waves diverge considerably from their  QC rendition. To recall, the QC begets a continuous branch of solutions with amplitudes ranging from zero to one and the corresponding velocities in the $[0,16/5]$ range, with one exceptional solution which attains the maximal velocity     $16/5$ and the maximal amplitude $2/\sqrt{3}$ and a much wider support. However, unlike the QC, the solution branch of the discrete antecedent bridges continuously  between the  disconnected QC solutions (see panels d and e in Fig~\ref{fig:bd_35f}). Furthermore, we also find an entire branch of discrete solutions, of which the QC is completely oblivious, with amplitudes hovering slightly below the exceptional QC amplitude
 $2/\sqrt{3}$ and the velocity $16/5$, with  widths which may be chosen at will (see panel  f in Fig~\ref{fig:bd_35f}). An enlarged display of the layer in the vicinity
 of point e in Fig.~\ref{fig:bd_35} is shown in Fig.~\ref{fig:bd_35p}. 
2
 Clearly, in this layer there is a very strong interaction between the non-linearity and the discreteness which has an essential impact on the resulting dynamics which the QC does not seem to be able to reproduce. Moreover, whereas in the G23 model both the discrete problem and its QC rendition yield kovatons residing on the singular manifold, as a flat-top solutions of arbitrary width, the presented almost flat-top solutions on the G35 lattice (though they appear to be {\it unstable}), occur only in the discrete model. Notably, their amplitude is close to $2/\sqrt{3}$, the maximal QC amplitude and their speed to a high accuracy is $25/8$ slightly below tha maximal QC speed.
 
    \item  As before, at large amplitudes there is another branch of discrete solutions (Fig~\ref{fig:bd_35f}, panels a,b,c) which may be approximated by the large amplitude QC solutions~\eqref{eq:p5c}. We were able to extend these solutions down to smaller amplitudes and velocities close to the c-labeled point(though their shape differs considerably from the simple cos form),  but failed beyond it  probably because, see  panel c in Fig~\ref{fig:bd_35f}, at the branch's edge the solution  becomes non-smooth.
    
     \item Similarly to the G23 case one notes the almost inverse relations between the norm (which up to a sign is wave's mass) and amplitude's response to changes in waves velocity. This comes out in Figs. \ref{fig:stpt35-1}-\ref{fig:coll35st2} where we display waves hoping from one branch to the other. The jump from right (left) branch to the left (right) results in amplitude increase (decrease), but as Fig 5 clearly shows, hardly in any changes in its mass.
\end{itemize}

\section{Interlaced Traveling Waves, ITW}
\label{sec:tptw}
We now proceed to unfold a more evolved class of solitary traveling waves which have distinct profiles at odd and even sites and appear as two interlaced solitary waves and may be considered as a simple form of moving breathers; they will be referred to as an interlaced traveling waves, ITWs. 
Denoting
\[
u_{2k}(t)=U\left(t-\frac{2k}{2\lambda}\right),\qquad u_{2k+1}(t)=V\left(t-\frac{2k+1}{2\lambda}\right)\;,
\]
begets a system of two coupled equations
\begin{align*}
\frac{dU}{dt}&=F[V(t-\lambda^{-1})]-F[V(t+\lambda^{-1})]\;,\\
\frac{dV}{dt}&=F(U(t-\lambda^{-1}))-F[U(t+\lambda^{-1})]\;.
\end{align*}
Setting $t=\lambda^{-1} s$ and following the same procedure  as in Sec.~\ref{sec:nie}, yields a system of two
coupled integral equations
\begin{equation}
\begin{aligned}
&\lambda U(s)+\int_{-1}^1 F[V(s+s')]\; ds'=0\;,\\
&\lambda V(s)+\int_{-1}^1 F[U(s+s')]\; ds'=0\;.
\end{aligned}
\label{eq:nint2}
\end{equation}

\subsection{Staggered Compactons}
\label{sec:sc}

In systems with  a $F(-x)=-F(x)$ symmetry, a simple interlaced 
solution may be derived from the basic TW. Setting  $V=-U$ reduces system~\eqref{eq:nint2} into one equation
\[
(-\lambda) U(s)+\int_{-1}^1 F[U(s+s')]\; ds'=0,
\]
which coincides with \eqref{eq:nint}. Thus any basic TW in a symmetric system produces an ITW with an opposite velocity.
Because the profiles at the odd and even sites are equal and opposite in sign, we shall refer to such wave as a "staggered compacton" in analogy with staggered solitons, cf.~\cite{Cai-Bishop-Gronbech-Jensen-94}.

\subsection{Interlaced traveling waves, ITW}
Applying the same numerical
procedure, including stability analysis, as in the basic TW case, we now proceed to unfold a more evolved branches of the interlaced traveling waves, 

\begin{figure}[!htb]
\includegraphics[width=\textwidth]{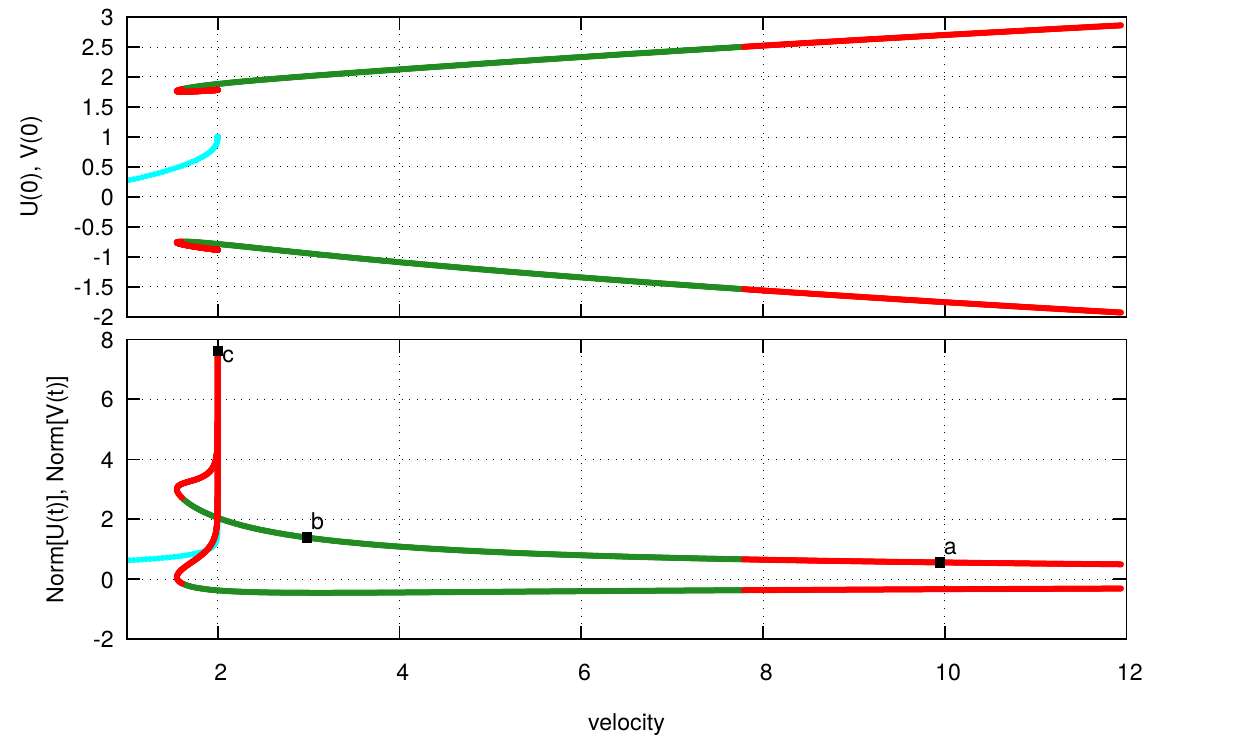}
\cprotect\caption{Branches of the  ITW in the G23 model \eqref{eq:gl23}. Green: stable waves, red: unstable waves. 
The basic TW branch is also shown (in cyan).}
\label{fig:bd_23p}
\end{figure}

\begin{figure}[!htb]
\includegraphics[width=\textwidth]{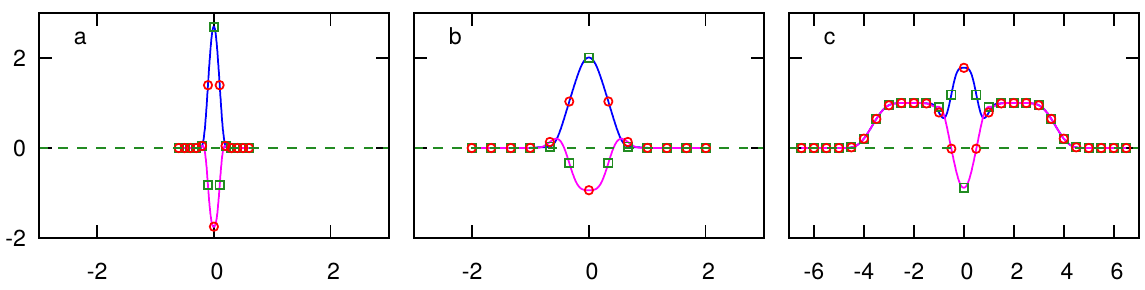}
\cprotect\caption{Display of the G23 ITWs  at the corresponding lettered points 
on the diagram in Fig.~\ref{fig:bd_23p}. Blue and magenta lines: profiles
at even and odd sites as functions of the time. Green squares and red circles:
snapshots on the lattice (when the center passes even and odd sites). Note  the more intriguing patterns in $a$ and $c$ which emerge in the vicinity of the QC critical speed.}
\label{fig:bd_23pf}
\end{figure}

{\bf The G23 model.}
Figures \ref{fig:bd_23p},\ref{fig:bd_23pf}
display the ITWs in the G23 lattice. 
Our understanding of these waves  is based on their large amplitude domain 
wherein the  waves have negative
velocities (thus propagate to the left) residing on a corresponding branch in Fig.~\ref{fig:bd_23}. In addition, since at large amplitudes the system is anti-symmetric in $u$, in this limit it also supports staggered compactons which propagate to the right. This corresponds to the ITW solution in  
Fig.~\ref{fig:bd_23p}, where at large positive velocities the solutions are seen to be nearly staggered (panel a in Fig.~\ref{fig:bd_23pf}). At smaller velocities/amplitudes, return of the quadratic  part in $F(u)$  to the game ruins the symmetry between odd and even sites (panel b in Fig.~\ref{fig:bd_23pf}). Formally,  since the staggering Ansatz leaves the $(-1)^k$ factor in front of the quadratic part,  its sign and thus its impact changes with the parity of $k$, causing the resulting  amplitudes of odd and even sites to be different. Notably, this branch of solutions extends
to velocities slightly under $2$, where it co-exists with the basic TW branch, also depicted in Fig.~\ref{fig:bd_23p}. There seems to be some sort of ``resonance'' between  these  waves which
induces large-norm
solutions at a velocity of $\approx 2$, with the resulting pattern having a remarkable flat-like appearance with a quasi-staggered compacton riding on its top
(panel c in Fig.~\ref{fig:bd_23pf}).

\begin{figure}[!htb]
\includegraphics[width=\textwidth]{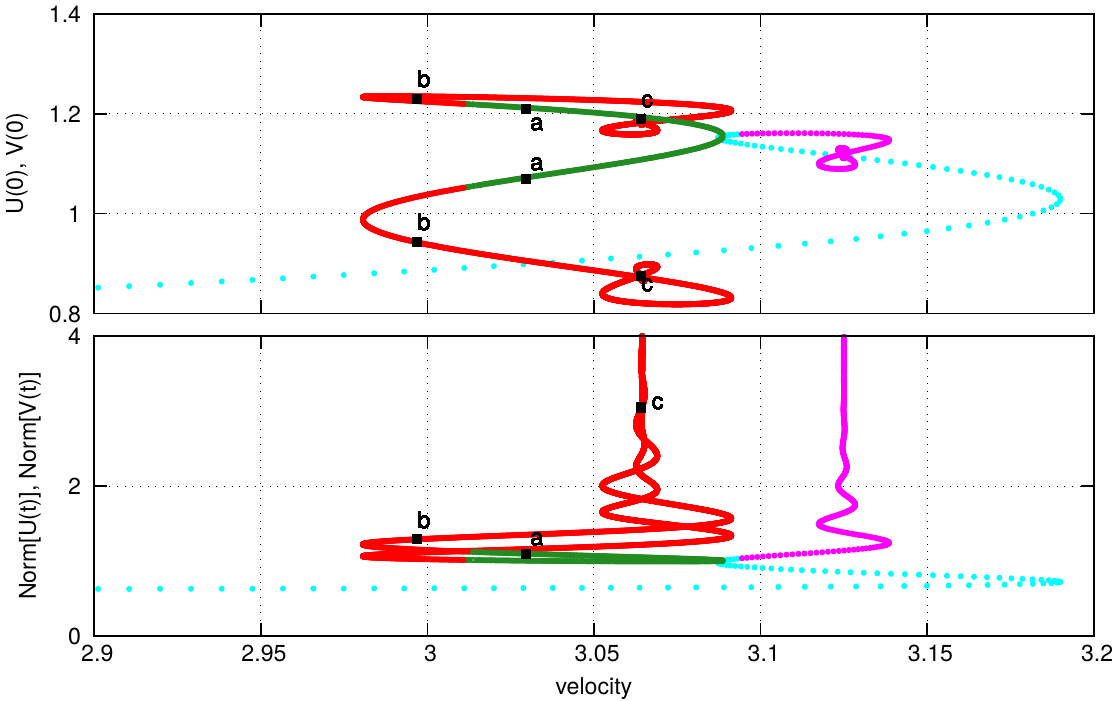}
\cprotect\caption{The interlaced TW branches of the G35 model. Note that the main action takes place in the vicinity of the critical transition. Green: stable waves, red: unstable waves. The presented waves do not seem to be related to the symmetrically  staggered compactons for, unlike the G23 model, the amplitudes of $U$ and $V$, being both are positive, are far from being reflection of each other, see fig~\ref{fig:bd_35pf}.  The basic TW branch is also displayed for comparison (stable TWs in cyan, unstable TWs in magenta).}
\label{fig:bd_35p}
\end{figure}

\begin{figure}[!htb]
\includegraphics[width=\textwidth]{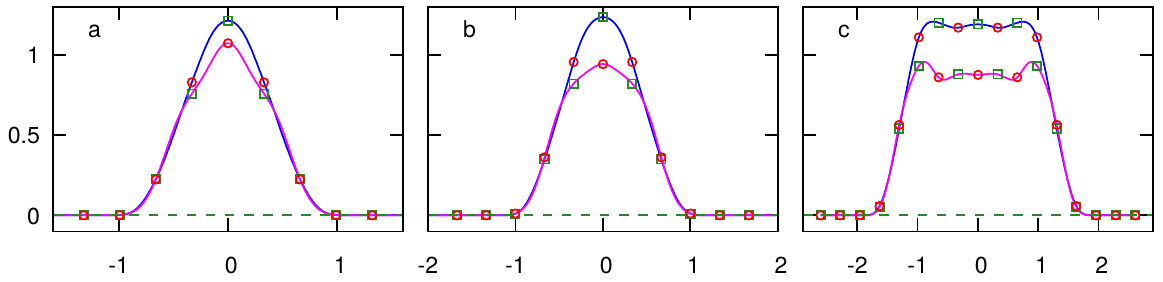}
\cprotect\caption{Profiles of the interlaced TW corresponding to the lettered points in Fig.~\ref{fig:bd_35p}. Blue and magenta lines: profiles
at even and odd sites, respectively, as functions of the time. Green squares and red circles:
snapshots on the lattice (when wave's center passes the even and the odd sites).}
\label{fig:bd_35pf}
\end{figure}

{\bf The G35 model.}
Figures~\ref{fig:bd_35p},\ref{fig:bd_35pf}  display the branches of the ITWs in the G35 lattice. The first thing to note is that the whole action takes place in a strip close to the critical speed $16/5$ where $F(u)$ is still positive. As already aforementioned few times, in this domain the discrete effects play an essential role.  However, this ITW branch does not appear to be a staggered version of the basic TW. Instead, it seems to be a result of a certain symmetry-breaking
in~\eqref{eq:nint2}. At its tip this branch touches the basic TW's branch. Close to the tip the profiles
at even and odd sites appear to be similar  (panel $a$ in Fig.~\ref{fig:bd_35pf}). Along the ITW branch the difference between the  even and odd sites profiles increases, but never becomes large: they look like mirror cases of the flat-top solitons; one amplitude is above $1$ whereas the other under it (panels b and c in Fig.~\ref{fig:bd_35pf}). Though it is tempting to relate the last pair with the basic TW in Fig.~\ref{fig:bd_35f},  the later attains a higher velocity on the threshold of the critical speed.

Finally, we  note that due to the anti-symmetry of the G35 lattice, any interlaced solution $(U,V,\lambda)$, like the ones in 
Figs.~\ref{fig:bd_35p},\ref{fig:bd_35pf}, 
always has an associated staggered counterpart  $(U,-V,-\lambda)$.

\section{Direct numerical simulations}
\label{sec:dns}

\begin{figure}[!htb]
\includegraphics[width=0.7\textwidth]{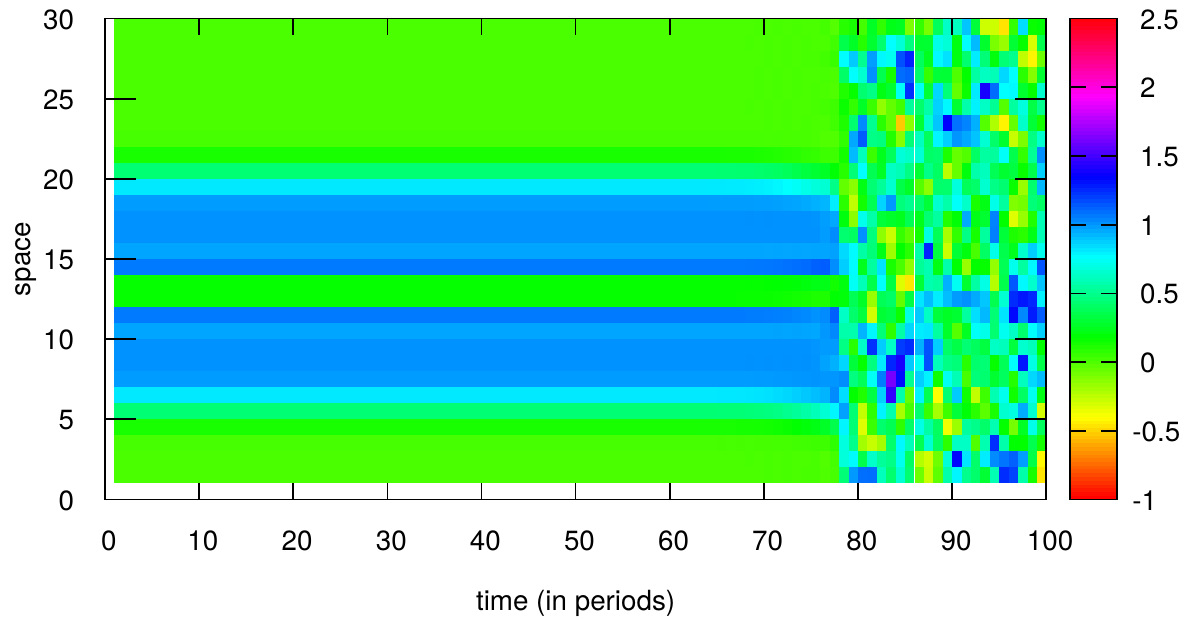}
\cprotect\caption{G23 lattice. Space-time diagram of the 
evolution of 
an interlaced TW 
(case c in Fig.~\ref{fig:bd_23pf}). It preserves its shape for about 75 periods(about 2250 sites).}
\label{fig:stabper23}
\end{figure}

Though in the numerical simulations of the lattice \eqref{eq:Fl}
we had no difficulty to observe  the unfolded TW yet, rephrasing Tolstoy's maxim,  whereas all stable waves are similar, each unstable wave is unstable in its own way, with this own way  being of great interest. Case in point;  though most of the unstable TW decompose within  few time units, in direct numerical simulations of the interlaced TW
of the G23 model (Fig.~\ref{fig:bd_23p}), several weakly unstable waves persisted for a long time. For instance, a wave with the maximal velocity of $\lambda\approx 12$ decomposes
only after  $t \approx 400$  traversing for about 4800 sites while preserving its shape.

The ITWs at the left side of the diagram in Fig.~\ref{fig:bd_23p}, with velocity 
$\lambda=2$
and large norms, are also relatively stable or, if you will, only weakly unstable. This is demonstrated in Fig.~\ref{fig:stabper23}, where  the waveform (displayed on a lattice of size $30$ at each period, $~$ $30/\lambda=15$)  remains intact for at least $75$ periods (i.e. up to $t \approx 1200$).

\subsection{Reversal of propagation direction}
\begin{figure}[!htb]
\centering
\includegraphics[width=0.8\textwidth]{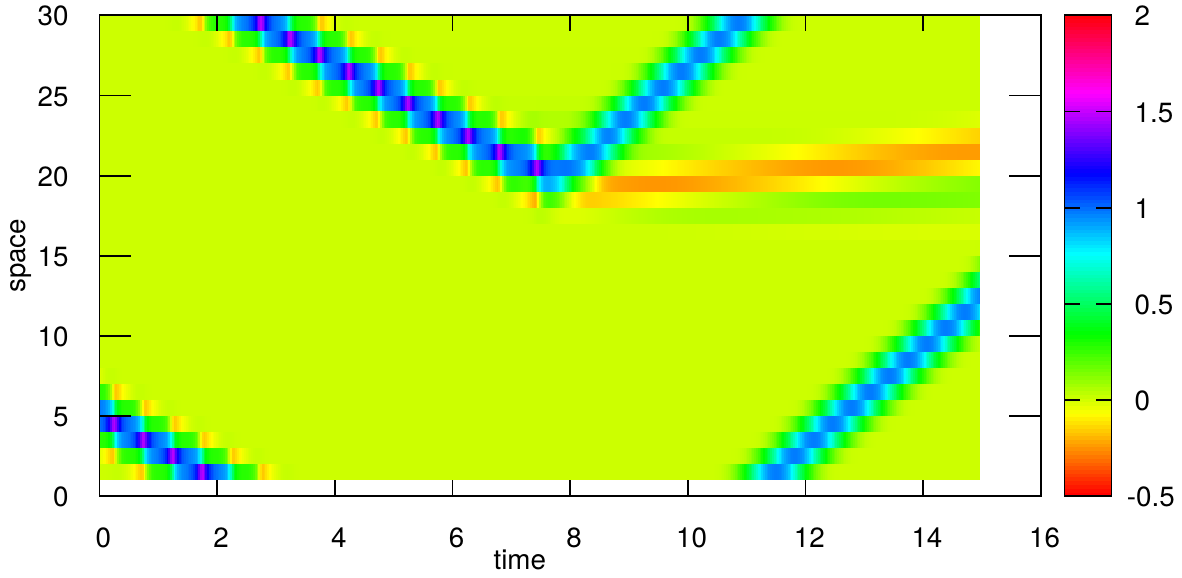}
\includegraphics[width=0.8\textwidth]{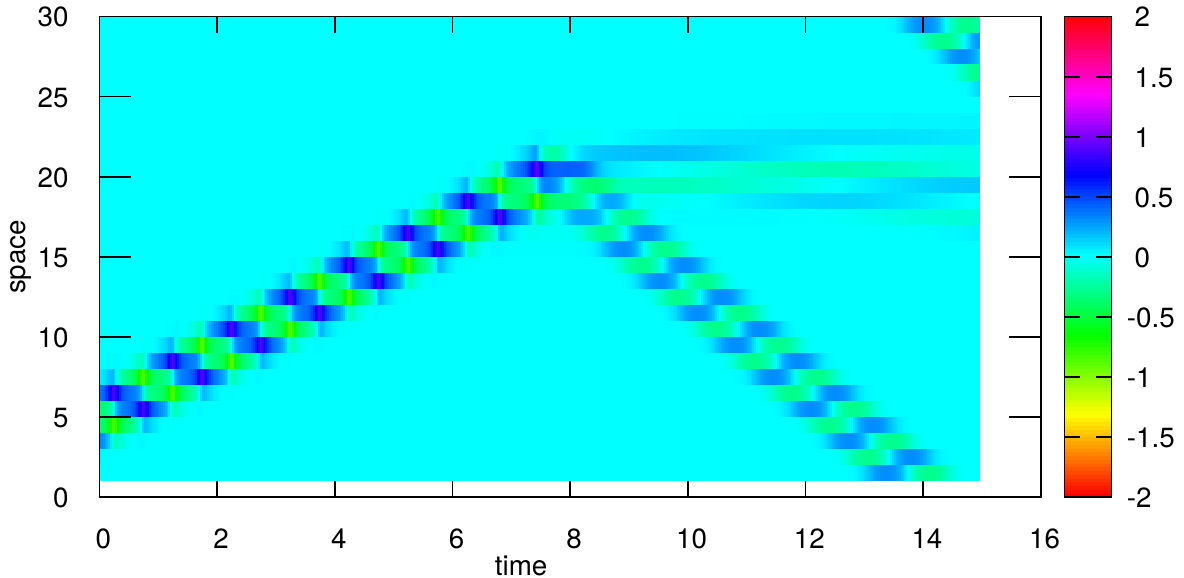}
\cprotect\caption{Reversal of soliton's path in the G35 model. Top panel: space-time diagram of a weakly unstable TW to the left. Due to its instability and proximity to branch's  edge, when wave's amplitude becomes too low to be sustained by the left branch, it hops to the other branch with a corresponding change of propagation direction. The lower panel displays the corresponding scenario for its  staggered twin with its propagation in the opposite direction. }
\label{fig:stpt35-1}
\end{figure}

A far more intriguing manifestation of the weak  instability in action emerges in the numerical simulations of TWs in the G35 lattice (Fig.~\ref{fig:bd_35}) where we have noticed  a weakly  unstable left-propagating wave,  $\lambda=-2.00467$, see  Fig.~\ref{fig:stpt35-1}, which propagates for about $8$ time units, comes to a halt, wobbles for a while, but then {\it reverses its direction turning into a stable right-propagating TW}. The same  phenomenon is observed in the corresponding staggered case (see the bottom panel of Fig.~\ref{fig:stpt35-1}). 

The key to the understanding of direction reversal is implicitly imprinted in Fig.~\ref{fig:bd_35} with its two panels complimenting each other. The upper panel displays an amplitude/velocity gap between the two branches wherein no propagation is admissible,  with the waves propagating in opposite directions on each side of the gap. However, the lower plate reveals that the amplitude/velocity jump is a non-event from the norm point of view,   for the norm hardly changes after the event (note the almost opposite amplitude-velocity and norm-velocity relations). The jump merely reshapes the wave's profile, and consequently its amplitude, with the resulting direction of propagation being a  byproduct of the branch on which it has landed.

\begin{figure}[!htb]
\centering
\includegraphics[width=0.8\textwidth]{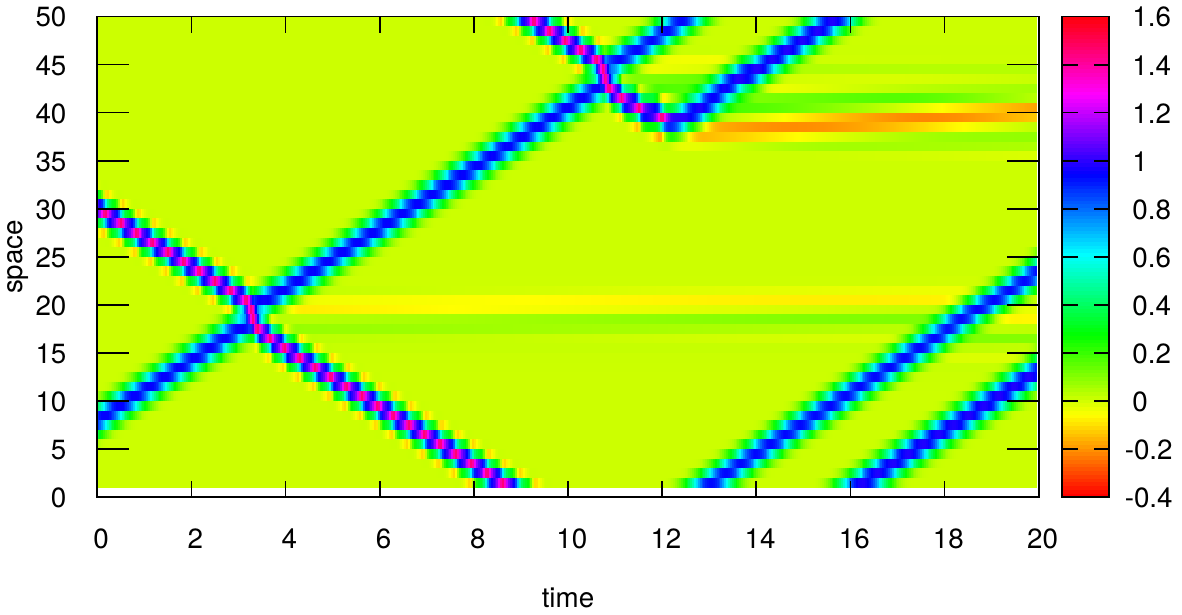}
\cprotect\caption{Collisions 
of a left-propagating (with amplitude $1.4024$) and right-propagating (amplitude $1.097$) solitary waves
in the G35 periodic lattice of length $N=50$. After two collisions the left-propagating
wave reverses its direction resulting in two right-propagating waves and some debris due to the collisions.}
\label{fig:coll35}
\end{figure}

\begin{figure}[!htb]
\centering
\includegraphics[width=0.8\textwidth]{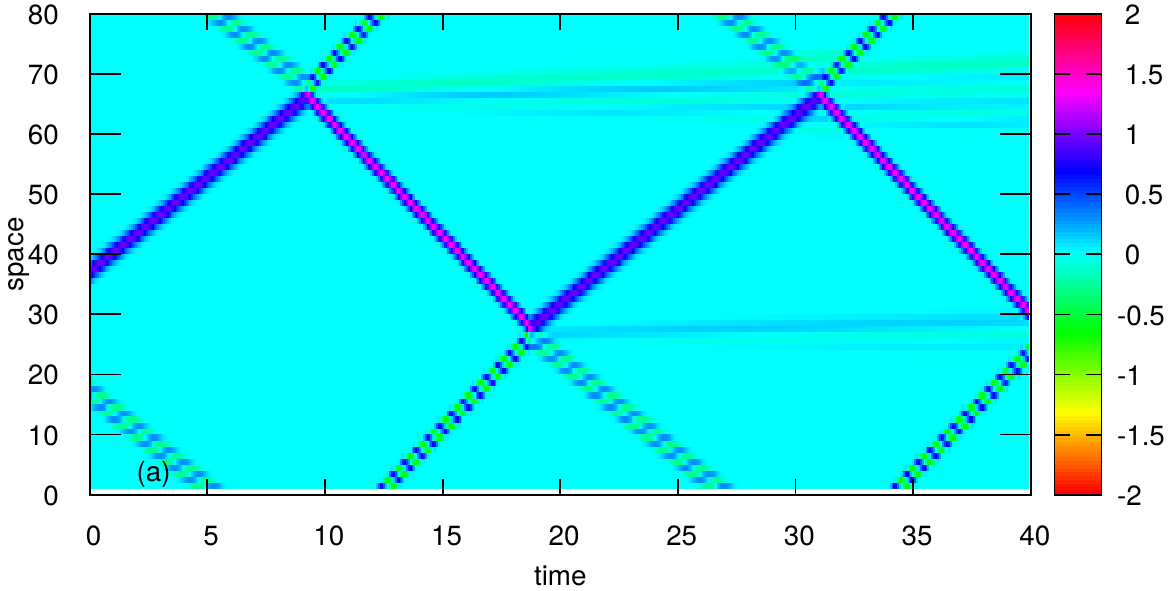}
\includegraphics[width=0.8\textwidth]{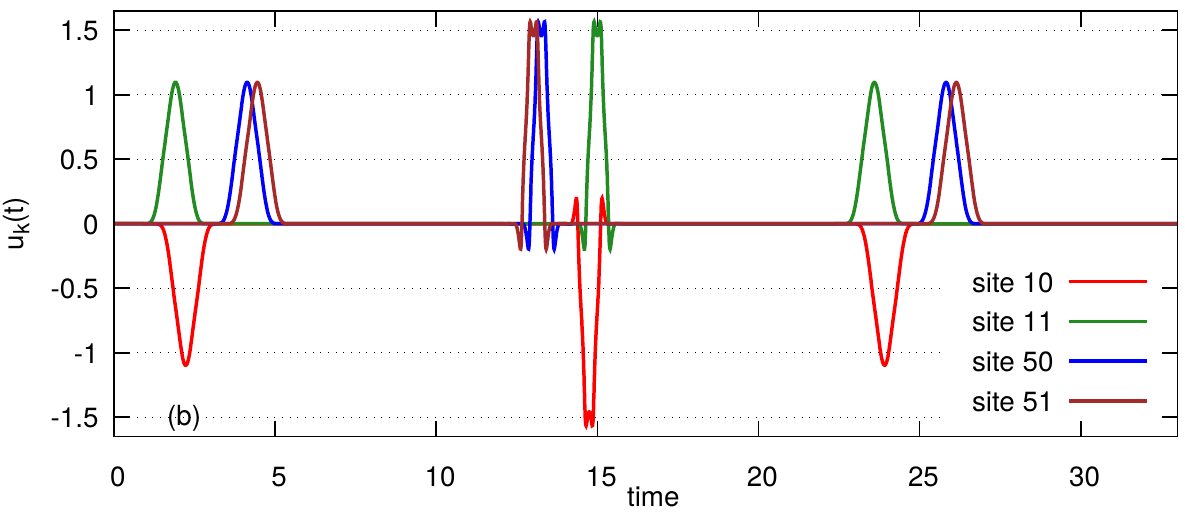}
\cprotect\caption{G35 periodic lattice of length $N=80$. Collisions 
of a  right-propagating wave (with amplitude $1.097$) with its staggered twin. 
(Note that the dynamics depends crucially on the initial distance between the two waves.  Shift by one lattice site shown in Fig.~\ref{fig:coll35st2} results in a very different dynamics). On  panel (a), a collision
at $t\approx 10$ transforms the right moving wave into a left-propagating wave on the left branch. With an opposite effect on its staggered twin. The
next collision at $t\approx 20$ brings them back to their original branch.
To visualize the "bouncing"  impact  on waves shape,
we display in panel (b) time series at two neighboring points pairs; $k=10$ (red) with $k=11$ (green) and $k=50$ (blue) with $k=51$ (brown). The red and green profiles 
at neighboring sites $k=10,11$ have opposite signs; as befits a staggered wave, whereas the blue and brown profiles at sites $k=50,51$, being merely shifted, corroborate a basic traveling wave. The chosen times cluster prior to the first, the second and the third collision. Though each encounter is accompanied by amplitude change - the mass (norm) itself hardly changes. 
}
\label{fig:coll35st}
\end{figure}

With this understanding we may readily follow Fig.~\ref{fig:coll35} where the two collisions between the right and the left moving waves suffices to relegate the left-moving wave into the other branch. Fig.~\ref{fig:coll35st} is a natural continuation of this scenario: here we have a  back and forth bouncing between the opposite branches, starting from the right branch  on Fig.~\ref{fig:bd_35}  and being pushed after one collision to the other branch. Following the second collision both waves  are back at their original, right, branch and the game continues. This could have gone indefinitely would it not be for the fact that the collisions are not entirely elastic with some debris created after each event.  

\begin{figure}[!htb]
\centering
\includegraphics[width=0.7\textwidth]{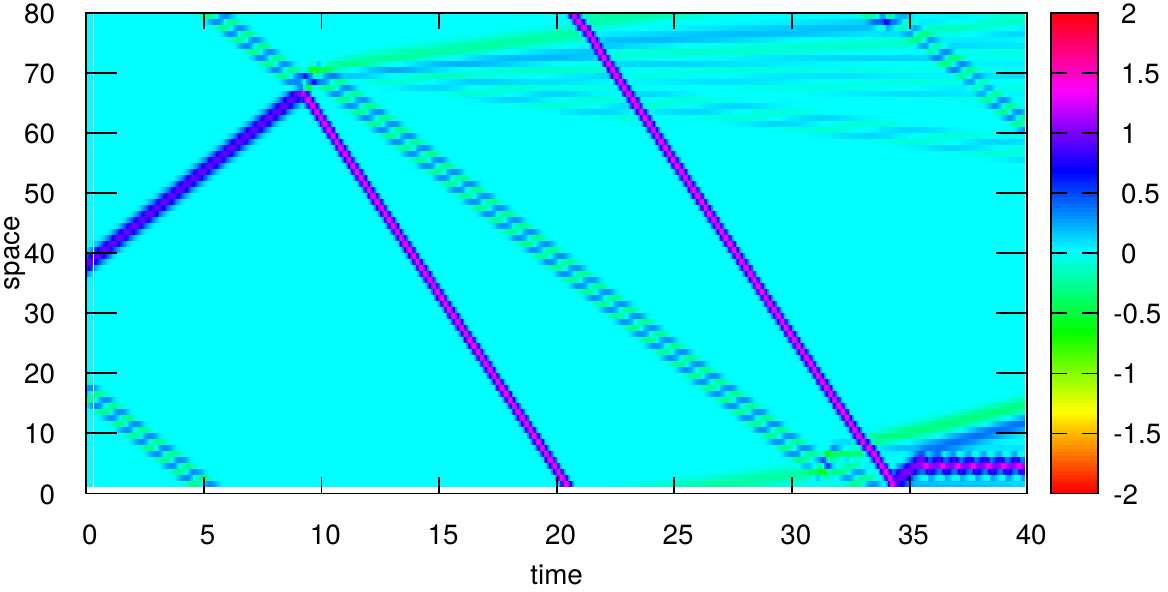}
\cprotect\caption{ Set-up similar to Fig.~\ref{fig:coll35st}, but with
the initial separation between the waves shifted by one site. Now after one collision the right-propagating wave turns into a fast left-propagating wave, but its staggered. left-propagating, twin stays the course and turns into another  left-propagating staggered wave.}
\label{fig:coll35st2}
\end{figure}

Unfortunately, as Fig.~\ref{fig:coll35st2} attests, the interaction {\it is very sensitive} to the collision phase of the TW and its staggered twin. A shift by merely one lattice site in the initial position of one of the waves causes the bouncing effect to disappear.

In the G23 model the story is much simpler;  no reversal of direction was ever observed. A weakly unstable TW located close to the right edge of the left branch, propagates for a while but then decomposes into disordered pieces with no observable coherent structure.

\section{Summary and Closing Comments}
\label{sec:con}

In the presented paper we have studied the traveling patterns of our strongly nonlinear
lattice versions of the Gardner-like  model equations G23 and G35, Eq.~\eqref{eq:gl23} and Eq.~\eqref{eq:gl35}, respectively. Three distinctive classes of traveling and non-traveling solitary patterns were unfolded

\noindent 1) Basic Traveling Waves,TW.\\
2) Interlaced Traveling Waves, ITW.\\
3) Stationary Solitary States.\\

Of the three classes, the second and the third have no quasi-continuum, QC, counter part, which is to say that space-wise they are essentially  discrete phenomena, unlike the first class which may be and was  very well replicated in part via the quasi-continuum.  Yet in spite of what may seem on its face as a limited success of the QC, it does play a vital role in delineating the road map without which most of the first and the second class solutions would not have been unfolded.

The utility of the QC extends well beyond its direct applicability, for though formally the original discrete system seem oblivious of the singularities, a hall mark of the QC, it appears that those singular manifolds are implicitly embedded in the system and both control and determine the crucial transition zone, though the PDEs which beget them are not valid. Consider, for instance, Figs.~\ref{fig:bd_23} and ~\ref{fig:bd_23p} of the G23 model and Figs.~\ref{fig:bd_35} and~\ref{fig:bd_35p} of G35 model, respectively. The role of the critical velocity $\lambda=2$ which emerges in the QC is clearly seen in Fig.~\ref{fig:bd_23}, but its role is even more impressive in Fig.~\ref{fig:bd_23p} that describes the quintessentially  discrete ITWs, which do not have a corresponding QC counterpart (and thus are unrelated to breathers found in the original Gardner equation), with the more interesting phenomena being clustered  around the critical values. The same effect is seen
in the G35 model with $\lambda=16/5$ emerging from the QC being the critical velocity. Besides its role as the edge of a ``QC'' branch in Fig.~\ref{fig:bd_35}, the more impressive impact is reserved for Fig.~\ref{fig:bd_35p} where all ITWs cluster in a narrow velocities strip located just under the critical velocity, whereas in the G23 model the ITWs in Fig.~\ref{fig:bd_23p} 'hug' the critical  
$\lambda= 2$ line from both sides, with a respective zoo of patterns, see Fig.~\ref{fig:bd_35pf}. All in all we have an amalgam of essentially discrete patterns concentrated around the macro constrains set by the QC.\\

As a motivation for the present work we have stated that the polynomial force was adopted because of the difficulty to treat the periodic case. Indeed, we were greatly helped by the existence of large amplitude regimes in the polynomial cases which helped us to unfold  new wave branches and new types of waves. Unfortunately, in the periodic cases  there are no large amplitude regimes and with everything being eternally coupled the presented results offer only a limited help.\\

The closing comment are reserved for the discrete stationary solutions briefly outlined in Sec.~\ref{sec:bm}. They span a finite plateau defined by the the roots of $F(u)=0$ and vanish initially  elsewhere (unlike the TWs and the ITWs which are associated with the roots of $F'(u)=0$). However, studies of their stability have revealed   a very sensitive dependence on their initial width, parity and lattice's width as well, which did not yield itself to a manageable  characterization. This topic is left for future studies. In passing we also note another class of kink-like excitations,  not discussed in the paper, for which the roots of $F''(u)=0$ play a key role. 

\acknowledgements
A.P. is supported in part by the Laboratory of Dynamical
Systems and Applications NRU HSE of the Ministry of Science
and Higher Education of Russian Federation (Grant No.
075-15-2019-1931). A. P. thanks A. Slunyaev for useful discussions.

\bibliography{gardbib}

\end{document}